\documentclass[twoside,reqno,a4paper]{article}
\usepackage{epsfig,cite}
\usepackage{amssymb,amsmath}
\usepackage{times}
\usepackage{fancyhdr}
\def\@fnsymbol#1{if case#1\hbox{}\or*\or\dagger\or\ddagger\or\mathcar''278\or\mathchar''27B\
or|\or**\or\dagger\dagger\or\ddagger\ddagger\else\@ctrerr\fi\relax}
\long\def\symbolfootnote[#1]#2{\begingroup%
\def\thefootnote{\fnsymbol{footnote}}\footnote[#1]{#2}\endgroup}

\newcommand{\runningheads}[2]{\markboth{{#2}}{{#1}}}
\newenvironment{Abstract}
{\bigskip\par\textsf{\textbf{Abstract}}\par}{}

\begin{document}

\sloppy \raggedbottom

\setcounter{page}{1}

\newpage
\setcounter{figure}{0}
\setcounter{equation}{0}
\setcounter{footnote}{0}
\setcounter{table}{0}
\setcounter{section}{0}
\LARGE\bfseries
\begin{flushleft}
A QMD description of the interaction of ion beams with matter\\
\end{flushleft}
\normalsize\normalfont
\begin{quote}
{\bf\underline{M.~V.~Garzelli}}$^{\, 1,2,\,}$\symbolfootnote[4]{$\,$ Corresponding author, {\it $\,\,$e-mail}: garzelli@mi.infn.it

$\,\,\,\,\,\,$ Proceedings 
of an invited talk presented at the 25$^{th}$ Workshop 
on Nuclear Theory, Rila Mountains,
Bulgaria, June 26 - July 1, 2006}, {\bf P.~R.~Sala}$^{\, 2}$, {\bf G.~Battistoni}$^{\, 2}$, {\bf F.~Cerutti}$^{\, 3}$,\\
{\bf A.~Ferrari}$^{\, 2,3}$, {\bf E.~Gadioli}$^{\, 1,2}$, 
{\bf F.~Ballarini}$^{\, 4}$, {\bf A.~Ottolenghi}$^{\, 4}$,\\
{\bf A.~Fass\`o}$^{\, 5}$, {\bf L.~S.~Pinsky}$^{\, 6}$, 
{\bf J.~Ranft}$^{\, 7}$\\    
\\
\small
$^{1}\mathrm{Dipartimento}$ di Fisica, Universit{\`a} di Milano, Milano, Italy\\
$^{2}\mathrm{INFN}$, Sezione di Milano, Milano, Italy\\
$^{3}\mathrm{CERN}$, Geneve, Switzerland\\
$^{4}\mathrm{INFN}$, 
Sezione di Pavia and Dipartimento di Fisica Nucleare e Teorica,
Universit{\`a} di Pavia, Pavia, Italy\\
$^{5}\mathrm{SLAC}$, Stanford, California, USA\\
$^{6}\mathrm{University}$ of Houston, Houston, Texas, USA\\
$^{7}\mathrm{Siegen}$ University, Siegen, Germany\\
\begin{Abstract}
Heavy-ion collisions can be simulated by means of comprehensive approaches,
to include the many different reaction mechanisms which may contribute. 
QMD models and their relativistic extensions
are examples of these approaches based on Monte Carlo techniques. 
In this paper are shown some results 
obtained by coupling a new QMD code, which
describes the fast stage of ion-ion collisions, 
to the e\-va\-po\-ra\-tion/fission/Fermi break-up and $\gamma$ de-excitation
routines present in the FLUKA multipurpose Monte Carlo transport 
and interaction code. In par\-ti\-cu\-lar, we compare the predicted 
neutron spectra 
to available experimental data from thin and thick target irradiations.
We show also some predictions of particle and charged fragment fluences
for the interaction of
C and Fe ions with a thick PMMA target, which may be useful to assess the
risk of side-effects in the hadron therapy of tumours.
\\
\end{Abstract}
\end{quote}

\section[]{Introduction}

Heavy-ion collisions at non-relativistic bombarding energies can 
be simulated by Quantum Molecular Dynamics
(QMD) calculations. 
These are comprehensive approaches which allow one, using Monte Carlo
techniques,
to take into account in a natural and straightforward 
way the whole of the different reaction mechanisms which contribute to a given
two-ion interaction as a function of the impact parameter and the bombarding 
energy. These features make them most suitable for describing processes 
where yields and fluences of emitted particles and
fragments have to be predicted and controlled, as it is needed in 
hadron therapy and radiation protection in space missions.  

Since nuclear fragmentation can be interpreted as the result
of nuclear density fluctuations, originated by the collisions,
a proper description of nucleon correlations is crucial for understanding 
this process. The QMD methods allow one to take into account
correlations in a natural and straightforward way. The time evolution 
of the projectile-target ion system in phase-space 
is calculated at each step of the simulation, by relating 
the spatial coordinates and momenta 
of each nucleon to the coordinates and momenta of all other particles.  
The nucleon wave-function evolution is evaluated 
by the Variational Principle, by 
mi\-ni\-mi\-zing the action corresponding to an Hamiltonian describing the
nucleon-nucleon interactions. 

In the original versions of QMD~\cite{aiche}, 
the nuclear wave-function is merely
given as a product of nucleon wave-functions. In more advanced versions,
e.g. the  Fermionic Molecular Dynamics (FMD)~\cite{fmd} 
and the Antisymmetrized Mo\-le\-cu\-lar Dynamics (AMD)~\cite{amd}, 
the fermionic nature of nucleons is 
properly taken into account. As far as we know, these last approaches
are used for investigating nuclear structure (e.g. for light 
exotic nuclei), but they have never been applied in a systematic way for
studying the interaction
of heavy-ion beams in thick targets, due to their complexity and the
required huge amount of CPU time. Since our aim is setting up an ion-ion
collision event generator to be used at
relatively low bombarding energies, from about one hundred 
up to a few hundred MeV/A, for describing the interaction and transport 
of heavy-ion beams in matter, considering thick composite targets of 
complex geometries, we choose to develop a non relativistic QMD code in 
which the fermionic nature of nucleons is taken 
into account in an approximate way. This is made using  
Pauli blocking factors in nucleon-nucleon scattering processes
and by giving an initial nucleon state distribution which forbids that
identical nucleons be in the same phase-space region~\cite{nuovavarenna}.

\runningheads{A QMD description of the interaction of ion beams with matter}
{M.V.~Garzelli {\it et al.}}

\section{The Hamiltonian}
The QMD calculations discussed in literature mainly differ in their  
Hamiltonian, both considering its terms 
and the strenght of the nucleon interaction coefficients.
In the following we briefly discuss, as reminded above, results which we have 
obtained using a non-relativistic Hamiltonian, i.e., assuming 
an istantaneous effective nucleon-nucleon interaction. 
Our Hamiltonian 
incorporates isospin and Coulomb effects, i.e.
{\it n}-{\it p}, {\it p}-{\it p}  and {\it n}-{\it n}
interactions have different strengths and radial dependence. 
Its nuclear part includes
an attractive Skyrme 2-body interaction
term and a repulsive Skyrme 3-body interaction term.
This one is crucial to reproduce the
saturation properties of nuclear matter at normal density ($\rho_0 \simeq
0.17\, \mathrm {fm}^{-3}$). 
A symmetry term takes into account isospin effects,
and a surface term, given by the sum of an attractive and
a repulsive term, is also included  to reproduce the decrease of
the nuclear potential at low $r$.
The nuclear terms of the Hamiltonian contain  parameters fitted to reproduce 
the observed properties of nuclear matter and finite nuclei. 

In order to compare most of the
calculated observables with experimental data, an accurate description of the 
de-excitation of residues produced after the fast nuclear interaction stage of
the reaction is also needed. In fact QMD 
can be used to study the fast overlapping stage of an ion-ion collision 
($\Delta t_{fast} \sim 10^{-22} s$), that leads
to the formation of pre-fragments, i.e. fragments which may be
excited. Other models, based on statistical considerations, 
are more suitable to describe the de-excitation of these fragments,
which may occur on a time scale se\-ve\-ral order of magnitudes
larger ($\Delta t_{de-ex}$ up to $\sim 10^{-15}$ s).
Thus, our QMD has been coupled to the de-excitation module available
in the FLUKA Monte Carlo transport 
and interaction code~\cite{flukacern, flukaweb}. 
At present, the de-excitation module allows one to take into account the
evaporation of light particles and intermediate mass fragments   
(up to A = 24), fission, Fermi break-up (in the
case of smaller fragments), and $\gamma$ emission. 

\section{Simulation of thin target experiments}

\begin{figure}[h!]
\includegraphics[bb=51 51 738 510, width=0.80\textwidth]{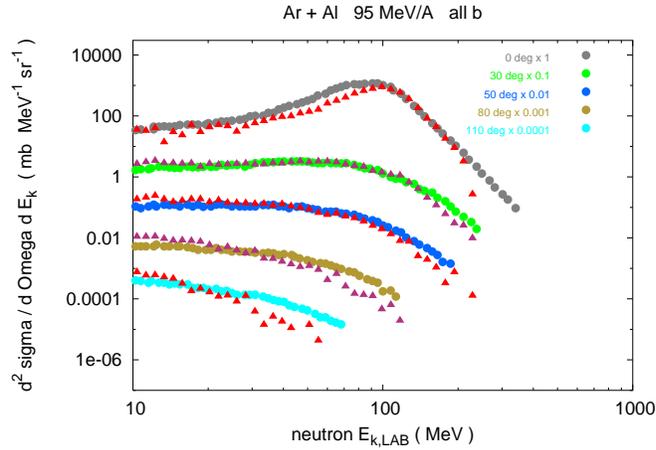}
\\
\caption{\small Double-differential spectra of the neutrons produced 
in the interaction of Ar and Al ions at 95 MeV/$A$ bombarding energy.
The theoretical distributions 
predicted by QMD~+~FLUKA (filled triangles)
are compared to the experimental data of Ref.~\cite{chiba} (filled
circles). }
\label{guraaralthin}
\end{figure}

To validate the low energy limit of our calculations we have analysed
neutron double differential spectra
in heavy-ion collisions at bombarding energies below 150 MeV/nucleon.
An example of the results obtained with our  QMD + FLUKA calculations
is shown in Fig.~\ref{guraaralthin} where our predictions are compared to the 
experimental spectra measured  by~\cite{chiba} in the interaction of 95 MeV/A 
Ar ions with a thin Al target. 
In this experiment the target thickness was chosen to ensure the
projectile ion energy loss be smaller than a few MeV. 
Our calculations (filled triangles) reproduce quite accurately the
 ex\-pe\-ri\-mental data (filled circles), 
especially at intermediate neutron emission angles 
(30 - 80 deg). 

\begin{figure}[h!]
\includegraphics[bb=51 51 738 510, width=0.80\textwidth]{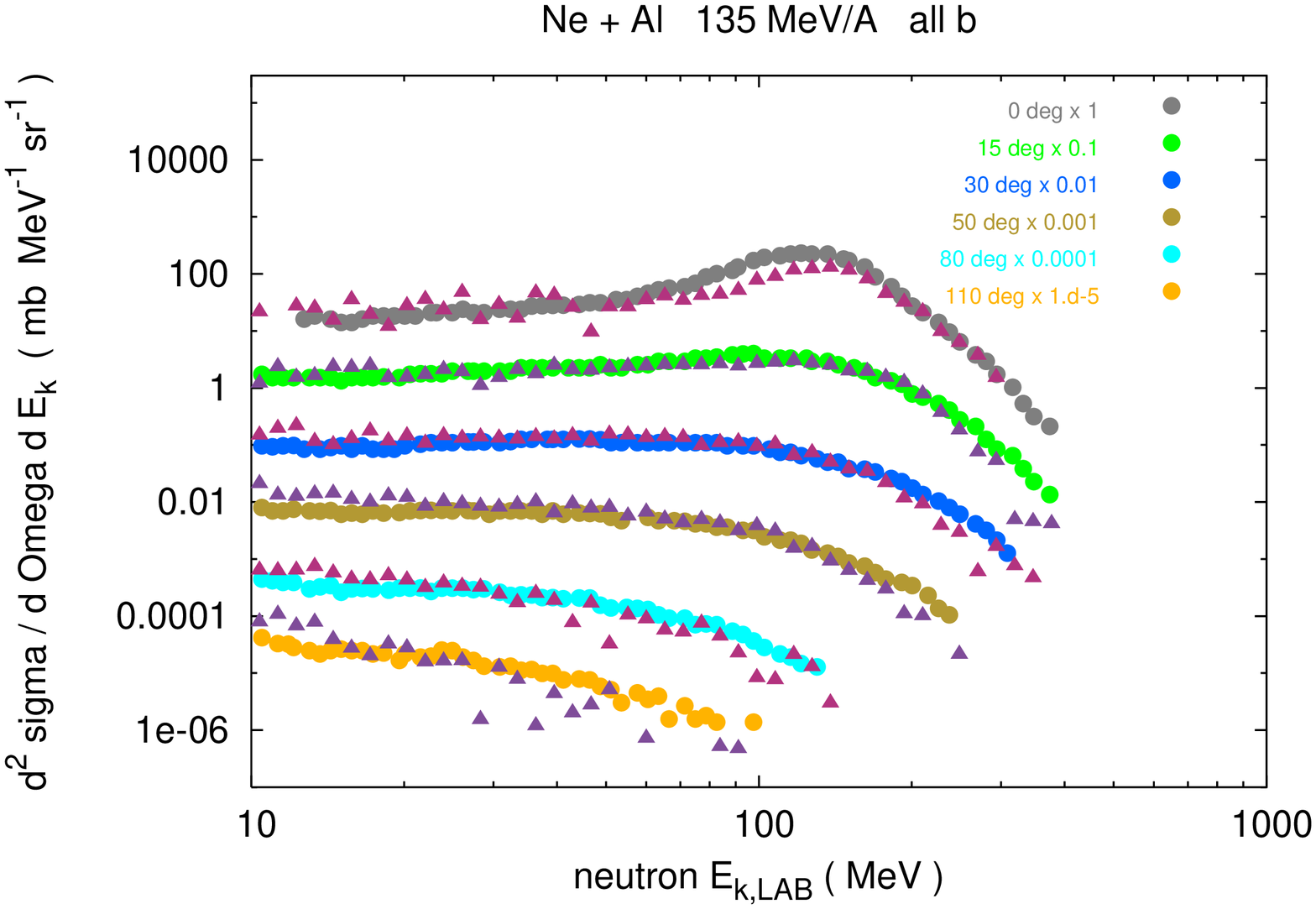}
\\
\caption{\small Double-differential spectra of the neutrons produced 
in the interaction of Ne and Al ions at 135 MeV/$A$ bombarding energy.
The theoretical distributions 
predicted by QMD~+~FLUKA (filled triangles) 
are compared to the experimental data of Ref.~\cite{chiba} (filled
circles). }
\label{guraneal}
\end{figure}

The QMD + FLUKA calculations reproduce with fair accuracy 
also the other experimental results given in Ref.~\cite{chiba}, 
as shown, e.g., in 
Fig.~\ref{guraneal} where the ex\-pe\-ri\-men\-tal 
and calculated spectra of the
neutrons emitted in the interaction of Ne ions with Al ions 
at 135 MeV/A bombarding energy are compared.

\begin{figure}[h!]
\includegraphics[bb=51 51 738 510, width=0.80\textwidth]{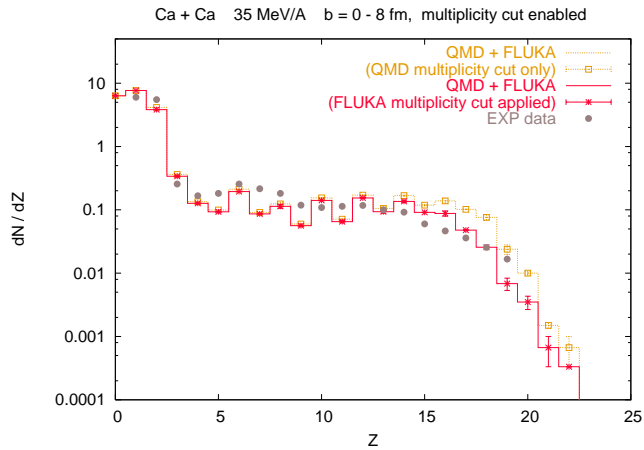}
\\
\caption{\small 
The theoretical prediction of the yield of the charged fragments
produced in the symmetrical Ca + Ca interaction at a 35 MeV/$A$ bombarding
energy (red filled circles and yellow squares, joined
by histograms) is compared to the experimental measurement for central 
collisions events (grey circles) made with 
the AMPHORA detector at SARA~\cite{ca35}. 
The reproduction of the data requires to simulate exactly the experimental set
up and contraints, as it may be appreciated by comparing the yellow histogram, 
obtained by imposing that the charged fragment multiplicity be more than 5
at the end of the nucleon interaction cascade described by QMD, 
to the red histogram, obtained by imposing that the charged fragment 
multiplicity be more than 10 at the end of the de-excitation stage, 
and fulfilling the requirement of quasicomplete 
events according to the detector acceptance, as in the experiment.}
\label{guraca35}
\end{figure}

An example of the ability of our calculations to reproduce the yield of the
emitted light and intermediate mass fragment is given 
in Fig.~\ref{guraca35}, for the reaction Ca + Ca at 35 MeV/A. To reproduce
accurately the data, the calculation must simulate accurately the experimental
constraints, such as those concerning the measured fragment multiplicity. 
The agreement of our calculations with the experimental data measured with the AMPHORA detector 
at SARA ~\cite{ca35}, is quite satisfactory.

\section{Thick target applications}
For applicative purposes one has often to deal with the interaction of an ion beam 
with
a thick target of complex geometry. In this case one must describe also the 
interactions of secondary particles and fragments with the target. To do this
our QMD code has been interfaced to the
FLUKA Monte Carlo transport and interaction code which  
allows one to study the transport of ions and secondary particles
in thick materials considering, in addition to nuclear interactions, 
many effects such as energy losses
due to medium ionization, bremsstrahlung, multiple scattering.
As reminded before, FLUKA also includes a nuclear de-excitation module, which
can be used to simulate the de-excitation of the hot fragments 
that may be present at the end of the fast stage 
of the ion-ion collisions described by the QMD calculations.

\begin{figure}[h!]
\includegraphics[bb=54 54 743 512, width=0.80\textwidth]{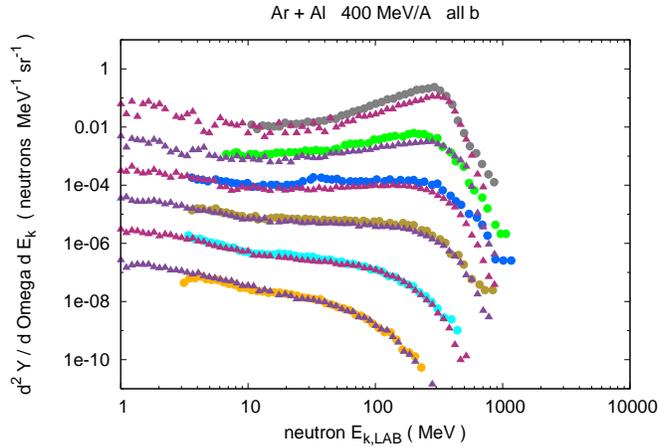}
\caption{\small Double-differential neutron spectra from the interaction of
Ar and Al ions at 400 MeV/A bombarding
energy. 
A 5.5 cm aluminium target 
was used to stop
completely the Ar incident beam.
The theoretical distributions
predicted by QMD~+~FLUKA (filled triangles)
at 0$^{\mathrm{o}}$, 7.5$^{\mathrm{o}}$, 15$^{\mathrm{o}}$, 30$^{\mathrm{o}}$, 60$^{\mathrm{o}}$, 90$^{\mathrm{o}}$  
emission angles
are compared to the experimental data of Ref.~\cite{giappospessi} (filled
circles).
}
\label{thickaral}
\end{figure}

An example of the ability of the QMD + FLUKA calculations to reproduce
neutron spectra observed in a thick target experiment is shown in 
Fig.~\ref{thickaral}, 
where our calculations are compared to the neutron spectra measured in 
the interaction of
an Ar beam with a thick Al target at 400 MeV/A bombarding
energy \cite{giappospessi}. 
The aluminium target used in the experiment
had a thickness $d = 5.5 \, 
{\mathrm {cm}}$, and was able to stop the incident beam. 
The agreement between the results of the calculated and
the experimental spectra is very encouraging, even considering the small
underestimation of the neutron yield along the beam direction. 
We emphasize that the reproduction of the absolute yield does not require any 
normalization coefficient. 
These calculations help to estimate the risk of side
effects in patient's treatment. One of the growing applications of FLUKA concerns just this 
field, considering also the biological damage to
the irradiated tissue~\cite{hadront1,hadront2,hadront3}.

%
\begin{figure}[p!]
\includegraphics[bb=0 21 498 498, width=0.80\textwidth, height=6cm]{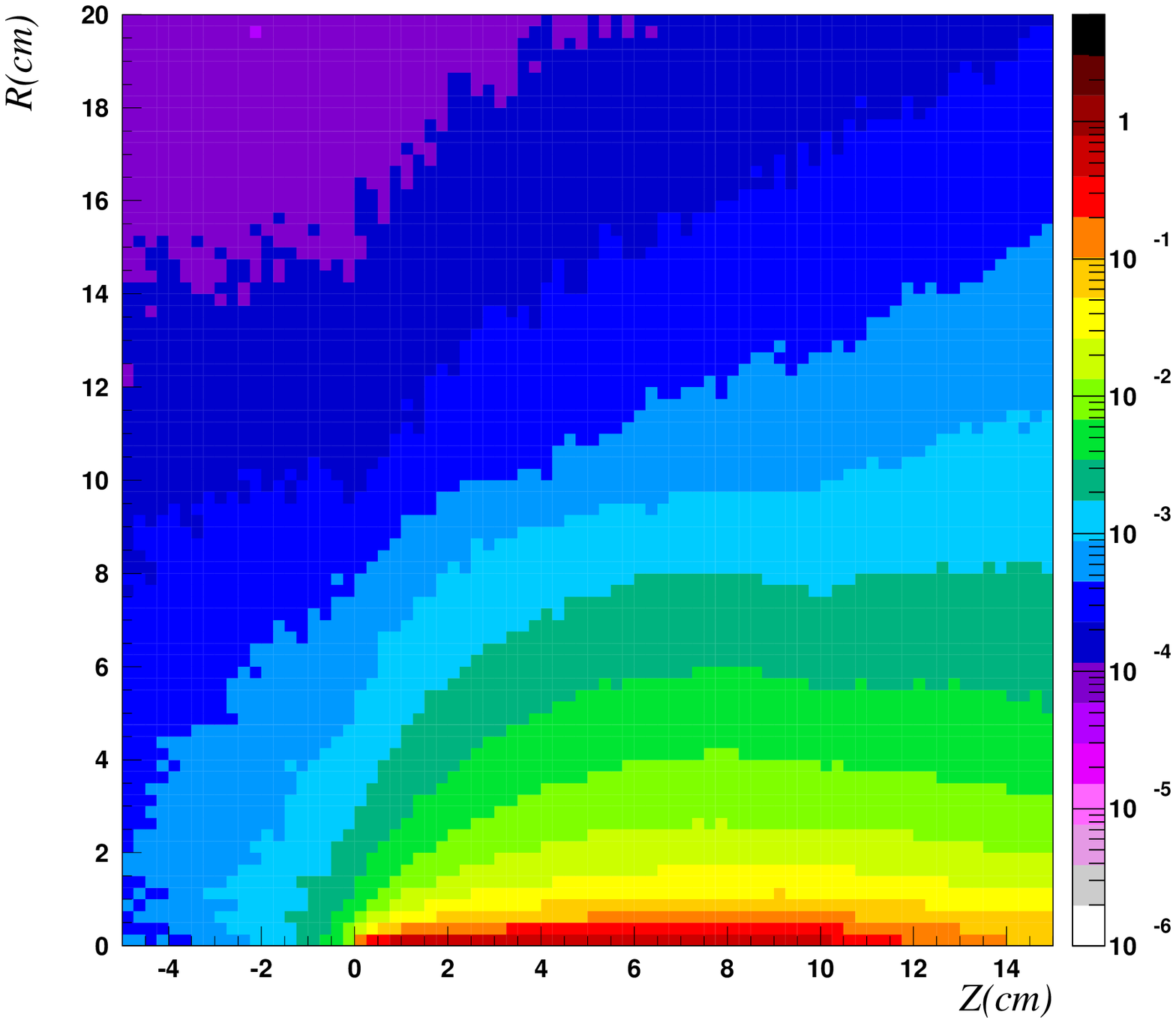}
\includegraphics[bb=0 21 498 498, width=0.80\textwidth, height=6cm]{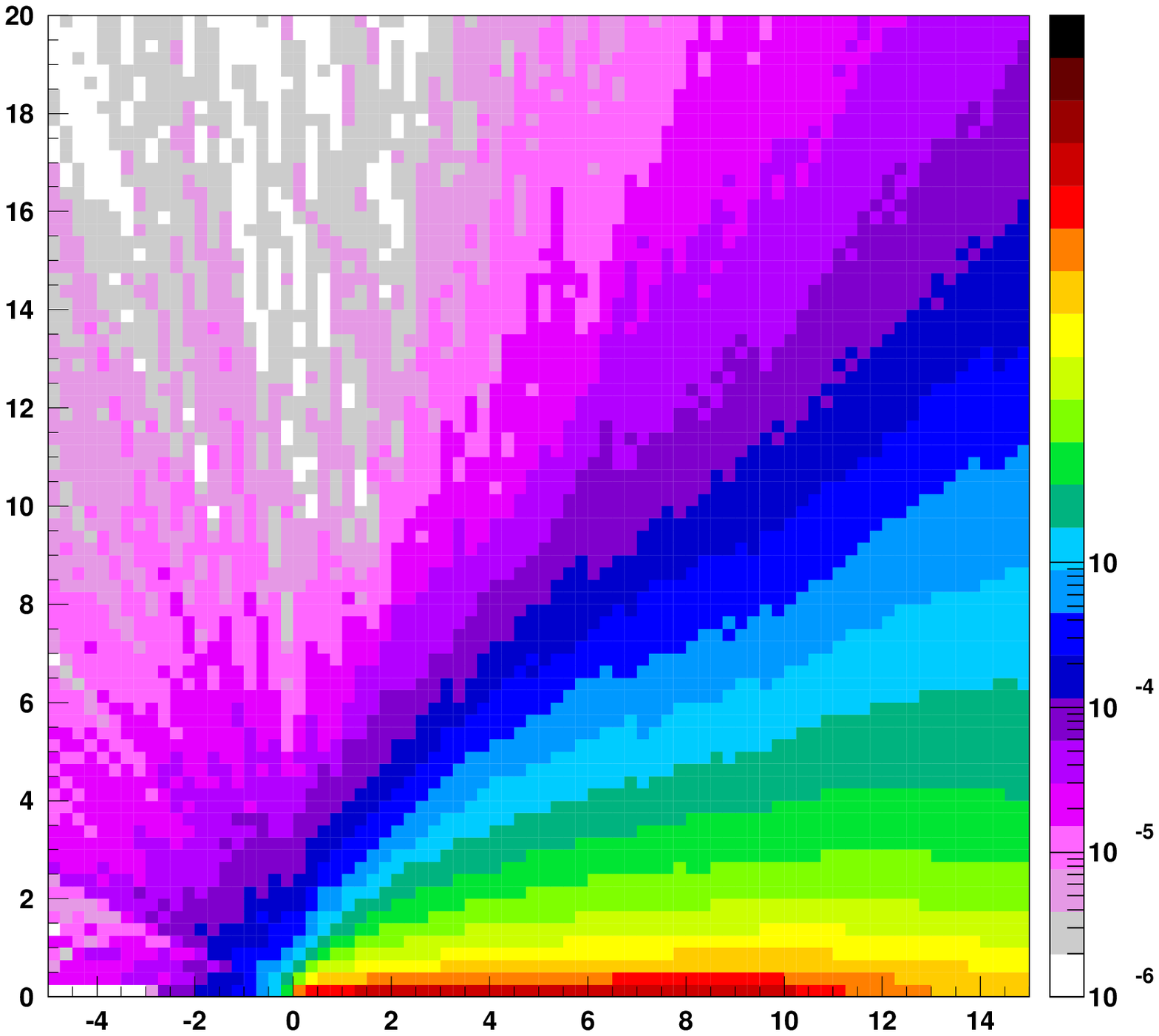}
\includegraphics[bb=0 21 498 498, width=0.80\textwidth, height=6cm]{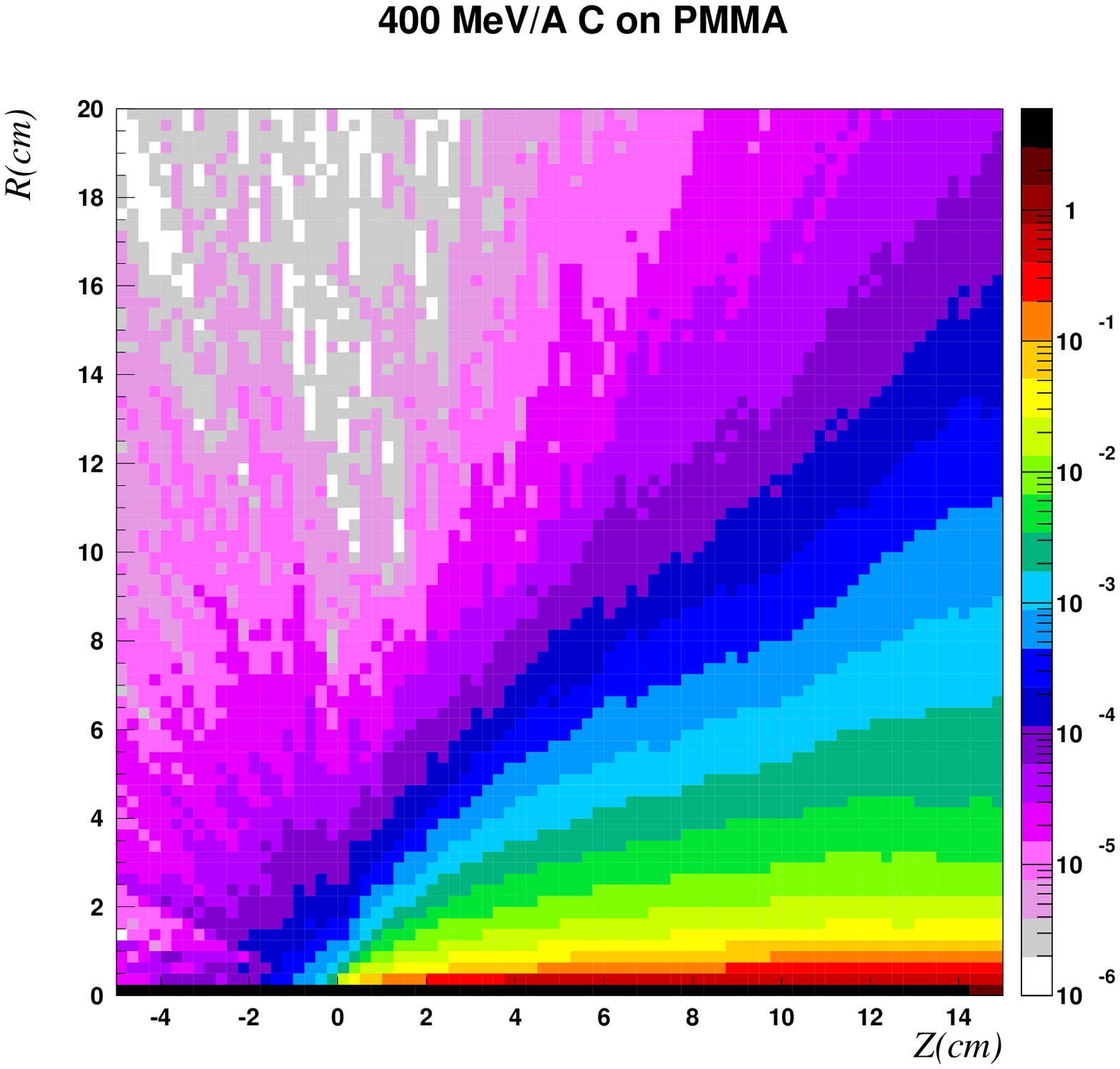}

\caption{\small Spatial distribution of the neutron ({\it top} panel), proton 
({\it intermediate} panel) and ion ({\it bottom} panel) fluences
for a  400 MeV/A 
C beam propagating along the axis of a 10 cm radius PMMA cylinder 
10 cm deep (which is not shown in the figure) in a cylindrical symmetry 
geometry. 
The plots show the results
of the simulations carried out with QMD + FLUKA. 
The cylinder front face is located at Z = 0 cm, while the beam is supposed
to come from the left along the Z axis (abscissa axis in the figure).
The PPMA cylinder is surrounded by air: the
formation of a cascade of particles and fragments also outside the cylinder
can be explained by the interactions of primary particles and secondary
products with the air surrounding the cilinder.}
\label{neutroncpmma}
\end{figure}
%
%

\begin{figure}[h!]
\includegraphics[bb=8 35 519 537, width=0.62\textwidth]{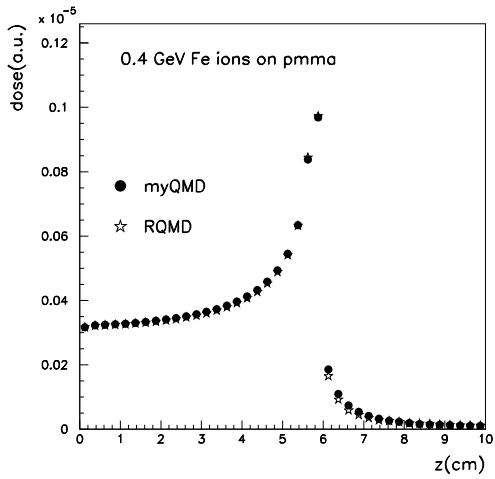}
\caption{\small 
Dose distribution (arbitrary unit) as a function of the propagation
depth 
for a Fe beam
propagating in the PMMA cylinder as in 
Fig.~\ref{neutroncpmma}, at 400 MeV/A bombarding energy. 
The cylinder front face is located
at Z = 0 cm (origin of the abscissa axis). 
The results of the simulation made with QMD + FLUKA (full circles)
are compared to those of the simulation with RQMD + FLUKA 
(empty stars). 
}
\label{bragg}
\end{figure}

The use of transport codes is necessary in hadron therapy studies for e\-va\-lua\-ting
the spatial distribution of the physical dose given to a patient. To do this one
must consider all primary and secondary particles, and their energy and angular
distribution as they propagate in the biological tissue.  
An example of the capabilities of the calculations in describing such processes 
is provided by the study of the propagation of a 400 MeV/A Carbon ion beam in
a 10 cm radius PMMA  cylinder 10 cm deep, surrounded by air. PMMA
(C$_{5}$H$_{8}$O$_{2}$) is a compound which allows to simulate  
the energy loss of a particle beam in biological tissues. 
The calculated spatial distribution of neutron, proton
and heavy fragment fluence 
(expressed in particles/$\mathrm{cm}^2$/primary ion)
are shown in the top, central and bottom panels of Fig.~\ref{neutroncpmma},
respectively. The comparison of the emitted particle's fluences  
shows that neutrons are more
spreaded out than protons and heavy-ions. Neutrons, originated
from primary and secondary  interactions of beam particles with the target, 
are not subject to Coulomb interaction and can thus propagate 
deep in matter. 
In Fig.~\ref{bragg} 
the dose distribution predicted
at dif\-fe\-rent depths inside the PMMA
cylinder previously considered is shown in the case of the bombardment with a 
400 MeV/A iron beam.  
In this figure, the results of the QMD + FLUKA calculations
are compared to those made with the  
current version of FLUKA using 
the Relativistic Quantum Molecular Dynamics code RQMD2.4~\cite{sorge, sorge2,
vecchiavarenna}.
The results of the two calculations practically coin\-ci\-de 
up to the Bragg peak
showing that the
prediction of the dose distribution  in this range region does not 
appear to be particularly sensitive
to the differences between QMD and RQMD2.4~\cite{nuovavarenna}.
On the other hand,
in the distal part of the
peak, slight differences concerning the total released dose appear.    
These are due to different implementations of nuclear interaction effects 
in the two codes, which are more pronounced at low interaction
energies.

\section{Conclusions and perspectives}
Fragmentation of ion beams propagating in matter can be simulated
by di\-na\-mi\-cal models,  considering nucleon correlations.
Quantum Molecular Dynamics models predict
nucleon-nucleon phase-space correlations in a straightforward
and natural way and allow one to describe the 
dynamical evolution of nuclear sy\-stems 
in the fast stage of heavy-ion
collisions. 
In this paper results 
are shown concerning particle and fragment
fluences and distributions, originated by the propagation and the
interactions of different beams
in low-Z targets. The si\-mu\-la\-tions have been performed with a QMD code
coupled to the FLUKA 
Monte Carlo general purpose
transport code. 

Especially gratifying is the reproduction of thick target data which are
si\-mu\-la\-ted 
by a fully microscopical calculation of the primary and secondary 
in\-te\-rac\-tions.  
The obtained results are encouraging in view of their use in hadron therapy 
and radioprotection in space missions.
For this purpose, further simulations and 
comparisons with experimental data 
to assess the ability of
the code in predicting heavy fragment distributions are at present under way. 
As shown in the
paper, charge fragment yields 
from symmetric central collisions 
are in very reasonable agreement with the experimental data.

\end{document}